\documentclass[12pt]{iopart}
\usepackage{graphicx}
\usepackage{multirow}
\usepackage{isotope}
\begin{document}

\title [QCD coupling constant ($\alpha_c$) and nonzero value of strange quark mass ($m_s\neq 0$)...]{QCD coupling constant ($\alpha_c$) and nonzero value of strange quark mass ($m_s\neq 0$) dependent stable stellar structure admitting observational results.}

\author{R Roy$^1$, K B Goswami$^2$, P K Chattopadhyay$^3$ and A Saha$^4$}
\address{$^{1,2,3}$ IUCAA Centre for Astronomy Research and Development (ICARD), Department of Physics, Cooch Behar Panchanan Barma University, Vivekananda Street, District: Cooch Behar, Pin: 736101, West Bengal, India}
\address{$^4$ Department of Physics, Alipurduar College, Alipurduar, Pin: 736122, West Bengal, India}
\eads{\mailto{royrohit477@gmail.com},\mailto{ koushik.kbg@gmail.com}, \mailto{pkc$_{-}$76@rediffmail.com}, \mailto{anirban.astro9@gmail.com} }


\begin{abstract}
This work discuses the effect of the QCD coupling constant ($\alpha_c$) on various physical properties of compact stars in the framework of the Tolman IV potential admitting the equation of state of MIT bag model with nonzero value of mass of strange quark mass ($m_s$). The internal matter, consisting of the deconfined phase of the $3$-flavour quark, is overall charge neutral due to the presence of electrons and is assumed to be strongly interacting. Interestingly, it is noted that the coupling constant $\alpha_c$ has an upper limit due to thermodynamic consistency and affects the stability of the stellar structure in terms of energy per baryon ($E_B$). Present model is suitable to study the properties of stars with mass of approximately $\leq 2.00~M_{\odot}$. The predicted radii of a few known stars from our model are in good agreement with the estimated values of radius obtained from the observations. Necessary energy conditions are obeyed inside the stellar configuration. Various stability conditions are carried out and it is found that within the range of parameter space used here the model is stable. 
\end{abstract}

%
\noindent{\it Keywords\/}:~{Compact object, Strange star, Quantum chromodynamics, Coupling parameter}
%
%
%
%

\section{Introduction}\label{s1}
Compact objects are of key importance in the study of astrophysics, nuclear and particle physics because of their high compactness and ultrahigh density ($\approx10^{15}~gm/cc$). This density regime initiates a series of processes that are beyond the knowledge of physicists to date. With the advances in theory, today, we have a better perception of structure and properties of high density matter inside compact objects. Depending on various parameters these objects are broadly classified as white dwarf, neutron star, strange star, gravastar and black holes. In case of white dwarf, the inwards gravitational pull is counter balanced by the outwards degeneracy pressure of electrons. On the other hand, neutron star is stabilised through the neutron degeneracy pressure. However, if the initial mass is more higher, neutron degeneracy pressure might not be able to support the inward gravitational pull. In this case, the volume of the star shrinks to comparatively a lower value. Thus, same mass may be accommodated in a lower radius or same radius may accommodate more mass. In both cases compatification factor (mass to radius ratio) and density increase. At such ultra-high density nucleons may be crushed into de-confined phase of quarks, forming a hybrid phase or pure quark phase. As per the Bodmer \cite{Bod} and Witten \cite{Wit} conjecture strange quark matter is the absolute ground state of hadrons instead of iron ($\isotope[56]{Fe}$), opens up a new area of study in high energy physics Based on this hypothesis compact objects of new features came in picture and are known as strange quark star (SQS) \cite{Bay, Alc, Gle}. As far as the stability of the system is concerned in this context, such stars may be composed of $3$-flavour quarks $u$, $d$ and $s$, instead of two flavour quarks, as the energy associated with per baryon is lower in case of the $3$-flavour quarks system ($u$, $d$ and $s$) than in the case of $2$-flavour quarks system, consists of $u$ and $d$. The possibility of charm ($c$) quark stars can be ruled out, as it is unstable against radial oscillations \cite{ket}.\par  
Recent mass measurements of $PSR~J1614-2230$ and $PSR~J0348+0432$ having mass $(1.97\pm0.04)M_\odot$ \cite{Pbd} and $(2.01\pm0.04)M_\odot$ \cite{Jan}, respectively indicate that the quarks strongly interact. This strongly interacting quark system is best studied under suitable application of Quantum ChromoDynamics (QCD). In the standard model formalism $SU(3)\times SU(2)\times U(1)$, QCD is one of the components also known as gauge theory which describes the strong interactions of quark colours and gluons. In the domain of high energy physics, it is well accepted that QCD is supposed to be fundamental theory used to describe the dynamics of strong interaction and in principle, it is possible that one could perform a detailed and comprehensive study on strange quark matter (SQM) by solving relevant equations of motion of quarks and gluons. The standard model relies heavily on the strong coupling parameter and quark masses for its validation and they are essential for testing its predictions. Precise measurements of masses of heavy-quark are essential for testing the  mechanism of possible mass generation proposed by Higgs \cite{GPL} and refining standard model parameter determinations \cite{JE1,JE2,JE3}. Similarly, accurate measurement of the strong coupling constant, $\alpha_c$, is critical for predicting branching ratios of Higgs boson \cite{GPL,SD} and evaluating the stability of the standard model vacuum \cite{DB,JRE}. In general, quark masses and $\alpha_c$ have been extracted by comparing perturbative QCD calculations with experimental data at high energy scales. These calculations involve expanding the QCD Lagrangian in powers of the strong coupling constant, $\alpha_c$, and then truncating the series at a certain order. The resulting expressions are then compared with experimental data to extract the values of the quark masses and $\alpha_c$. The accuracy of these extractions depends on the order of $ \alpha_c$ of the perturbative expansion, the quality of the experimental data, and the theoretical uncertainties associated with the calculations. For example, $\alpha_c$ can be extracted from $e^+e^-$ annihilation \cite{GD1,GD2,GA,SB,RAD,RA2,TG2,AH}, from jets \cite{VMA,BM,SC,VK1,VK2}, $\tau$-decay \cite{PAB,AP,MD,DB2}, or deep inelastic scattering \cite{JB,SA,PJD,ADM,LAH,RDB}. However, the strength of the interaction between quarks remains unknown, as QCD is intractable in the nonperturbative regime. Therefore, phenomenological models play important roles in extracting and determining the relevant properties of strongly interacting quark matter. Many theoretical models are based on QCD approach, such as the model of Nambu-Jona-Lasinio \cite{YNam,TXia}, the the quasiparticle model \cite{VG,LJL}, the mass-density-dependent model \cite{XJW}, global colour symmetry model \cite{WB} and the MIT bag model \cite{far}. Among these MIT bag model \cite{far} is the simplest and is useful for complex quantum many body systems. The interaction within the soup of quarks is labelled with a parameter $\alpha_c$, which is the called strong interaction coupling parameter and is related to the QCD coupling constant $g$ as $\alpha_c=\frac{g^2}{4\pi}$ \cite{WMYAO}. In this paper, we focus on how equilibrium conditions are modified with $\alpha_c$ and the strange quark mass ($m_s$).\par 
According to Chandrashekhar, the maximum mass for white dwarf is $1.4$ $M_{\odot}$ \cite{srini,sc} and it was recently modified to be $2.58$ $M_{\odot}$ \cite{DM} considering one Landau energy level. However, for neutron star (NS) or strange quark star (SQS), the upper limit of the maximum mass is still unknown. It is widely believed that the maximum mass of NS is very much model dependent or specifically depends on Equation of State (EoS). Softening or stiffening of the EoS has some effect on the maximum mass. In the articles \cite{Alf,Foz}, strong interactions between the quarks may lead to the stiffening of EoS, whereas those with hyperon and kaon condensation or with a transition from pure NS to a hybrid or pure SQS may soften the EoS. The possible detection of the gravitational wave events by the LIGO and VIRGO interferometers \cite{Bpa}, further constrains the EoS. Therefore, choosing a suitable EoS is of key importance for studying the internal structure of compact objects. Many authors have used liner EoS \cite{Bor,Hae}, quadratic EoS \cite{MMa}, polytropic EoS \cite{GNSGC} and chaplygine type EoS \cite{Pb} to study the internal properties of compact stars. In the articles \cite{JS,Pau,Lug,Chow} successful applications of MIT bag model EoS in the frame work of general relativity have been studied. Goswami \etal \cite{Kau}, used the modified MIT bag model EoS considering finite value of strange quark mass ($m_s$). Strange quark matter can be modelled as a Fermi gas composed of $u$, $d$ and $s$ quarks and electrons ($e^{-}$) so that the system is overall charge neutral. In the case of MIT bag model, the quark confinement is characterised by a constant term defined as the bag constant $B_g$. $B_g$ is actually the difference of the energy between perturbative and nonperturbative vacua \cite{Pbh}. Additionally, this phenomenological parameter is determined by the underlying strong interaction dynamics. We consider the first order effects of $\alpha_c$ in our model, to study how the physical parameters are affected. It is important to note that strange quark mass ($m_s$) and coupling constant $\alpha_c$ are important in determining the relationship between the energy density ($\rho$) and baryon number density ($n_A$), which in turn determines the binding energy ($E_B$) of the star. Furthermore, the small electron abundance depends sensitively on $m_s$ and $\alpha_c$. As neutrino cooling depends on the presence of electrons, the neutrino emissivity depends on $\alpha_c$ and $m_s$. In short, the relevant properties of strange matter can be determined using the physical quantities $B_g$, $\alpha_c$, and $m_s$. The main aim of this paper is to construct a suitable stellar model to describe strange quark stars in the framework of the Tolman IV potential.\par  
According to Ruderman \cite{Ruderman}, highly dense compact objects, where the density surpasses that of nuclear matter, may exhibit anisotropic matter composition. Within these objects, the existence of type 3A superfluids could contribute to the development of anisotropy within the fluid sphere. The inclusion of pressure anisotropy in the study of the relevant properties of matter at extremely high densities has also been considered by several authors \cite{BRL,VV,FR1,FR2,FR3,MKALAM,HOSSEIN,MKALAM2,ASAHA}. The possible origins of anisotropy are superfluidity \cite{RA}, the existence of a strong magnetic field \cite{FW}, slow rotational motion \cite{Her} etc. Anisotropy may have some non negligible effects on different parameters of the star, such as the stiffness of the EoS, maximum mass and radius. Many authors \cite{KA,Bvi,FES} have studied the effects of anisotropy on the physical properties of many spherically symmetric ultra dense fluid spheres. As an application of our model, we predict the radii of various compact stars and study the effects of $\alpha_c$ on various physical parameters with $m_s\neq{0}$. \par
The paper is organised as follows: In \sref{s2}, we have studied the thermodynamics of the SQM and constraints on $B_g$ due to non zero coupling between the quarks of the $3$-flavour quark system. In \sref{s3}, solutions for the EFE are obtained using Tolamn-IV type metric ansatz, and various boundary constraints on the system are imposed. The restriction on the coupling constant ($\alpha_c$) is discussed in \sref{s4}. The maximum mass and radius from the present model are given in \sref{s5}. \Sref{s6} includes some physical applications of our model. In \sref{s7}, various stability conditions are studied and finally in \sref{s8} we conclude the outcomes of our model.\\

\section{Thermodynamics inside the star}\label{s2}
The thermodynamic equilibrium of a mixture of massless $u$ and $d$ quarks, electrons ($e^{-}$) and strange quarks ($s$) with finite mass ($m_s$) allows for the transformation mediated by the weak interactions between quarks and leptons. Now quarks are spin-$\frac{1}{2}$ fermions carrying a fractional electric charge and gluons are massless spin-$1$ bosons carrying a colour charge. The quarks interact with each other, and this interaction is mediated by the gluons. The strength of this interaction is represented by the coupling parameter $\alpha_c$ \cite{XYW}. For a fixed quark mass, $\alpha_c$ is the only free parameter of the Lagrangian of interaction in the QCD \cite{HJ}. Additionally, it is one of the three fundamental coupling parameters in the standard model of particle physics. The value of $\alpha_c$ depends on the energy scale or momentum transfer ($Q$) of the process \cite{XIAO}. Owing to asymptotic freedom, which is a property of the QCD, $\alpha_c$ is small at high energies or short distances. In this domain of the QCD, the perturbation method is a reliable tool for calculating the thermodynamic potential. The precise value of $\alpha_c$ is a crucial parameter in the study of QCD \cite{ADeur}. Previously, many authors have considered a constant value of $\alpha_c$ \cite{KS,XJW1,far}. We also adopt a constant value of $\alpha_c$. We consider corrections due to $\alpha_c$ upto first order only to reduce the complexities and to illustrate the effects of $\alpha_c$ and to study the necessary modifications if any in the structure of strange quark stars. we focus on the simpler case of first-order corrections only. We start with the perturbative expansion of the thermodynamic potential density in presence of cold quark matter composed of $3$-flavour quarks. The total thermodynamic potential ($\Omega$) up to the first order in $\alpha_c$ \cite{Alc,far} is expressed as:

\begin{equation}
	\Omega=~\Omega_u+\Omega_d+\Omega_s+\Omega_e,\label{1}
\end{equation}
where, $\Omega_i$, ($i= u,~d,~s,~e^{-}$) represents the thermodynamic potentials for the respective particles. The contributions to the total thermodynamic potential from massless $u$ and $d$ quarks are as follows:
\begin{equation}
	\Omega_u=-\frac{\mu_u^4}{4\pi^2}\left(1-\frac{2\alpha_c}{\pi}\right),\label{2}
\end{equation}
\begin{equation}
	\Omega_d=-\frac{\mu_d^4}{4\pi^2}\left(1-\frac{2\alpha_c}{\pi}\right),\label{3}
\end{equation}
whereas, the contribution from the strange quark of nonzero mass is,
\begin{eqnarray}
	\Omega_s&=&-\frac{1}{4\pi^2}\left[\mu_{s}\nu_{s}(\mu_s^{2}-\frac{5m_s^2}{2})+\frac{3m_s^4}{2}\ln{\frac{\mu_s+\nu_s}{m_s}}\right. \nonumber \\
	&& \left. -\frac{2\alpha_c}{\pi}\left\{3\left(\mu_s\nu_s-m_s^2\ln{\frac{\mu_s+\nu_s}{\mu_s}}\right)^{2}-2\nu_s^{4}\right. \right. \nonumber \\
	&& \left.\left. -3m_s^{4}\ln^{2}{\frac{m_s}{\mu_s}} +6\ln{\frac{\epsilon_R}{\mu_s}} \left(\mu_s\nu_{s}m_s^2-m_s^{4}\ln{\frac{\mu_s+\nu_s}{m_s}}\right)\right\}\right].\label{4} 
\end{eqnarray}
where, $\nu_s=(\mu_s^2-m_s^2)^{1/2}$, $\epsilon_R$ is termed as the renormalization point, and $\mu_i$ represents the chemical potential of the $i^{th}$ particle, where $i=u,d,s$ and $e^{-}$. As the electron does not participate in the strong interaction, its contribution to the thermodynamic potential is,
\begin{equation}
	\Omega_e=-\frac{\mu_e^4}{12\pi^2}.\label{5}
\end{equation}
The renormalization point ($\epsilon_R$) must be chosen carefully. If we incorporate all the orders of $\alpha_c$ in our calculations then the physical parameters are independent of $\epsilon_R$. However, as we include correction up to the first order in $\alpha_c$, there is some dependence of the physical parameters associated with the stars on $\epsilon_R$ which are not allowed physically. Such dependence may lead to some results that are not viable. To minimize such dependence, one must choose the the value of $\epsilon_R$ properly. For this reason, we consider $\epsilon_R=313~MeV$ \cite{far}, as this value is close to natural energy scale and comparable with the average chemical potential. However, Duncan \etal \cite{Dsw} has made the choice $\epsilon_R=m_s$, which may lead to strong dependence on $\epsilon_R$. Here, we consider deconfined quark matter which is locally charge neutral and in $\beta$-equilibrium. Chemical equilibrium is reached via the following weak interactions,
\begin{eqnarray}
	d\rightarrow u+e+\bar{\nu_e},\;\;\;\;\;\;s\rightarrow u+e+\bar{\nu_e} \nonumber \\
	u+e \rightarrow d+\nu_e,\;\;\;\;\;\; u+e \rightarrow s+\nu_e, \nonumber \\
	s+u \rightleftharpoons d+u\label{6}
\end{eqnarray}
Equation \eref{6}, leads to the following relations:
\begin{eqnarray}
	\mu_d=\mu_u+\mu_e,\;\;\;\;\;\;\mu_s=\mu_u+\mu_e, \nonumber \\
		\mu_s=\mu_d\equiv\mu.\label{7}
\end{eqnarray}
The contribution of the neutrinos can be neglected because of the larger mean free path they escape from the system. Local charge neutrality depends on the presence of electrons and leads to the following relation:
\begin{equation}
	\sum_{i=u,d,s,e^{-}}  n_iq_i=0,\label{8} 
\end{equation}
where $q_i$ and $n_i$ represent the charge and number density of the $i^{th}$ type particle and is given by $n_i=-\frac{d\Omega_i}{d\mu_i}$. The pressure $(p)$, energy density $(\rho)$ and the total baryon number density $(n_A)$ of the deconfined quark assembly are evaluated from the relations given below:
\begin{equation}
	p=-\sum_{i=u,d,s,e^{-}}\Omega_i - B_g,\label{9}
\end{equation}
\begin{equation}
	\rho=\sum_{i=u,d,s,e^{-}}\left(\Omega_i+\mu_in_i\right)+B_g,\label{10}
\end{equation}
\begin{equation}
	n_A=\frac{1}{3}(n_u+n_d+n_s).\label{11}
\end{equation}
The expression connecting the relation between pressure and density are also known as EoS of the matter, can be obtained using equations \eref{9} and \eref{10} in presence of nonzero mass of strange quark and coupling constant ($\alpha_c$) as: 
\begin{equation}
	\rho=3p+4B_g+4\Omega_s+n_s\mu.\label{12}
\end{equation}
 
Since we have only two independent chemical potentials $\mu$ and $\mu_e$, the simultaneous solution of equations \eref{8} and \eref{9} at the surface of the star gives the values of the chemical potentials ($\mu$, $\mu_e$) for a given parametric choice of $\alpha_c$, $m_s$ and $B_g$.

\begin{figure}[ht]
	\begin{center}
		\includegraphics[width=10cm]{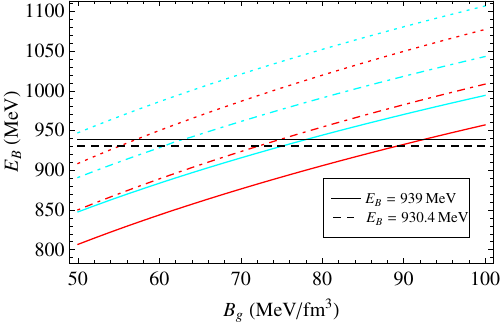}
		\caption{Energy per baryon plot for different $m_s$ and $\alpha_c$. The red and cyan colours indicate $m_s=~50$ and $150~MeV$, respectively. The solid, dotdashed and dotted lines indicate $\alpha_c=~0.0,0.3$ and $0.6$ respectively.}\label{fig1}
	\end{center}
\end{figure}

As suggested by Goswami \etal \cite{Kau}, there are three stability windows depending on the value of energy per baryon, namely: $(1)$ the stable region $(E_B<930.4~MeV)$, $(2)$ the metastable region $(930.4~ MeV<E_B\leq939~MeV)$ and $(3)$ the unstable region $(E_B> 939~MeV)$ where $930.4~MeV$ is the typical binding energy of iron and $939~MeV$ is that of neutrons \cite{MAD}. We have plotted the variation of $E_B$ with $B_g$, as shown in \fref{fig1}, for different parametric choices of $m_s$ and $\alpha_c$. It is observed that stability window decreases with increasing $\alpha_c$ and also decrease with increasing $m_s$ for fixed $\alpha_c$. All the results are tabulated in \tref{tab1}. Notably, the stability window changes in the presence of $\alpha_c$ and $m_s$.
\begin{figure}[ht]
	\begin{center}
		\includegraphics[width=10cm]{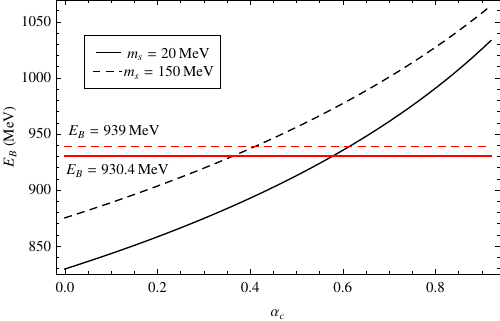}
		\caption{Variation of $E_B$ with $\alpha_c$ for different choices of $m_s$. Here the black solid and dashed lines indicate the variations corresponding to $m_s=20~MeV$ and $150~MeV$ respectively. The region below the red solid line represents the stable window for quark matter. The region between the red solid and dotted lines represent the metastable window for quark matter. On the other hand, above the red dotted line, the region corresponds to unstable quark matter.}\label{fig2}
	\end{center}
\end{figure}
In \fref{fig2}, the dependence of $E_B$ on $\alpha_c$ is shown for two different parametric choices of $m_s$ with $B_g=57.55~MeV/fm^3$. This plot shows that the value of $E_B$ increases with increasing $\alpha_c$ and moves from the stable to the unstable region. Additionally, with increasing $m_s$, $E_B$ moves towards the unstable region for a smaller value of $\alpha_c$.  
\begin{figure}[ht]
	\begin{center}
		\includegraphics[width=15cm]{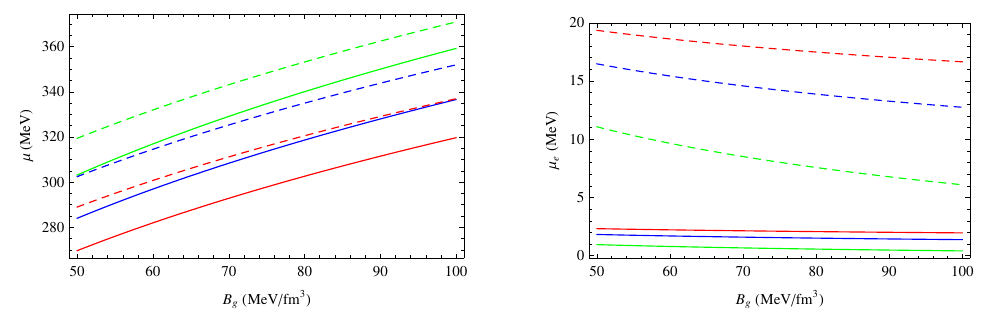}
		\caption{Variation of chemical potential with $B_g$. In the left panel the chemical potentials of the quarks and in the right panel those of the electrons are shown. Red, blue and green colours correspond to $\alpha_c=0,~0.3$ and $0.6$, respectively. The solid and dashed lines represent the variations for $m_s=50$ and $150~MeV$, respectively.}\label{fig3}
	\end{center}
\end{figure}

\Fref{fig3} shows the variation of the chemical potential with $B_g$ for quarks (left panel) and electrons (right panel) for the parametric choices of $m_s$ and $\alpha_c$, respectively and indicates that for a fixed $B_g$ and $m_s$, $\mu$ increases with increasing $\alpha_c$, whereas $\mu_e$ decreases with increasing $\alpha_c$. Additionally, with a higher $m_s$ value, this dependence is stronger. For a given $m_s$, as the coupling between the quarks increases, the number of electrons necessary to maintain the charge neutrality condition in the system decreases.  

\begin{table}[ht]
\centering
\caption{\label{tab1}Constraint on $B_g$ in $MeV/fm^3$ for different stability windows at zero external pressure.}
\resizebox{0.8\textwidth}{!}{$
	\begin{tabular}{@{}c|c|ccc}
		\hline
		$\alpha_c$             & $m_s$    &   Stable             & Metastable                 & Unstable \\
		                       & $MeV$    & ($E_B\leq{930.4}~MeV$) & ($930.4<E_B\leq{939}~MeV$)        & ($E_B>939~MeV$)\\ \hline
		 \multirow{3}{*} {0.0} & 50       & $B_g<88.90$          & $88.90<B_g<92.36$          & $B_g>92.36$\\
		                       & 100      & $B_g<82.66$          & $82.66<B_g<86.11$          & $B_g>86.11$\\
		                       & 150      & $B_g<74.81$          & $74.81<B_g<77.74$          & $B_g>77.74$\\ \hline      
		 \multirow{3}{*} {0.3} & 50       & $B_g<72.02$          & $72.02<B_g<74.81$          & $B_g>74.81$\\
		                       & 100      & $B_g<66.97$          & $66.97<B_g<69.76$          & $B_g>69.76$\\
		                       & 150      & $B_g<60.32$          & $60.32<B_g<62.98$          & $B_g>62.98$\\ \hline 
		 \multirow{3}{*} {0.35} & 50       & $B_g<69.33$          & $69.33<B_g<71.94$          & $B_g>71.94$\\
		                       & 100      & $B_g<64.52$          & $64.52<B_g<67.10$          & $B_g>67.10$\\
		                       & 150      & $B_g<58.20$          & $58.20<B_g<60.62$          & $B_g>60.62$\\ \hline       
    \end{tabular}$}
\end{table}

\section{Field equations and their solution}\label{s3}
In case of a spherically symmetric  and static fluid sphere the interior space-time is defined by the following line element,
\begin{equation}
	ds^2=-e^{\nu(r)}dt^2+e^{\lambda(r)}dr^2+r^2(d{\theta^2}+\sin^2{\theta}d{\phi^2}),\label{13}
\end{equation}
where $e^{\nu(r)}$ and $e^{\lambda(r)}$ are the unknown metric potentials and are functions of $r$ only. The Einstein Field Equation (EFE), which connects the geometry of space and matter is given by: 
\begin{equation}
	R_{\mu\nu}-\frac{1}{2}g_{\mu\nu}R=\frac{8{\pi}G}{c^4}T_{\mu\nu},\label{14}
\end{equation}
where $R_{\mu\nu}$, $g_{\mu\nu}$, $R$ and $G$ are known respectively as the Ricci tensor, metric tensor, Ricci scalar and the gravitational constant of Newton. The energy momentum tensor $T_{\mu\nu}$ in case of anisotropic distribution of fluid is given as:
\begin{equation}
	T_{\mu\nu}=\mbox{diag}~(-\rho,p_r,p_t,p_t),\label{15} 
	\end {equation}
where $\rho$ is the energy density, $p_r$ and $p_t$ represent respectively the radial and tangential pressure. Combining equations \eref{13}, \eref{14} and \eref{15}, we derive the equations for $\rho$, $p_r$ and $p_t$ as given below:
\begin{equation}
		\rho=\frac{\lambda^{\prime}e^{-\lambda}}{r}+\frac{1-e^{-\lambda}}{r^2},\label{16}
\end{equation}
\begin{equation}
		p_r=\frac{\nu^{\prime}e^{-\lambda}}{r}-\frac{1-e^{-\lambda}}{r^2},\label{17}
\end{equation}
\begin{equation}
		p_t=e^{-\lambda}\left[\frac{\nu^{\prime\prime}}{2}+\frac{{\nu^{\prime}}^2}{4}+\frac{\nu^{\prime}-\lambda^{\prime}}{2r}-\frac{\nu^{\prime}\lambda^{\prime}}{4}\right],\label{18}
\end{equation}
where we considered $8{\pi}G=1,c=1$ and the prime denotes differentiation with respect to $r$. In anisotropic formalism, the pressure anisotropy is defined as $\Delta=(p_t-p_r)$. Introducing the Durgapal-Bannerji transformation \cite{Durgapal} as,
\begin{equation}
	x=r^2\;\;\;\;\;Z(x)=e^{-\lambda(r)}\;\;\;\;\;y^{2}(x)=e^{\nu(r)},\label{19}
\end{equation}
equations \eref{16}-\eref{18} are modified in terms of the new variable $x$ and we obtain the following:
\begin{equation}
	\rho=-2Z_x+\frac{1-Z}{x},\label{20}
\end{equation}
\begin{equation}
	p_r=4Z\frac{y_x}{y}-\frac{1-Z}{x},\label{21}
\end{equation}
\begin{equation}
	p_t=Z\left[4\frac{y_x}{y}+4x\frac{y_{xx}}{y}+\frac{Z_x}{Z}+2x\frac{y_x}{y}\frac{Z_x}{Z}\right],\label{22}
\end{equation}
where the subscripts '$x$' and '$xx$' refer to the derivatives of first and second order of $y$ and $Z$ with respect to $x$, respectively. To derive an exact solution, we use equation \eref{12} along with equations \eref{20} and \eref{21} and obtain the following equation:
\begin{equation}
	Z_x+C_{1}Z=C_{2},\label{23}
\end{equation} 
where, $C_{1}=2Z(\frac{1}{x}+3\frac{y_x}{y})$ and $C_{2}=2(\frac{1}{x}-B_g-\Omega_s-\frac{n_s\mu}{4})$. Equation \eref{23} is a first order differential equation and can be solved if we choose a suitable form of $y$, which is singularity free and satisfies all the general physical conditions. In this paper we use the Tolman-IV \cite{tolman} type ansatz for $y$ as:
\begin{equation}
	y^2=A(1+ax),\label{24}
\end{equation}
where $A$ and $a$ are two unknown constants to be determined. This form of the metric ansatz is well behaved, continuous and nonsingular in the interior of the star and therefore in constructing a stellar model such form can be acceptable. Substituting equation \eref{24} into equation \eref{23}, we obtain the following solution:
\begin{eqnarray}
	Z&=&-\frac{1}{30(1+ax)^3}\left[-30+20B_{1}x+15ax(3B_{1}x-4)\right. \nonumber \\
	&& \left. +9a^2x^2(4B_{1}x-5)+2a^3x^3(5B_{1}x-6)\right],\label{25}
\end{eqnarray}
where, $B_{1}=B_g+\Omega_s+\frac{n_s\mu}{4}$. The integration constant is taken to be zero to avoid geometrical singularity at the centre. The expressions for the physical parameter such as $\rho$, $p_r$, $p_t$ and anisotropy ($\Delta$) in this model are as follows:
\begin{eqnarray}
	\rho=\frac{1}{10(1+ax)^4}\Big[20B_{1}+a\Big(30+55B_{1}x+ax\Big(45+69B_{1}x \nonumber  \\
	\;\;\;\;\;\;+ax(27+42B_{1}x+2ax(3+5B_{1}))\Big)\Big)\Big],\label{26}
\end{eqnarray}
\begin{eqnarray}
	p_r=\frac{1}{30(1+ax)^4}\Big[-20B_{1}+a\Big(30-105B_{1}x+ax(45-171B_{1}x\nonumber \\ 
	\;\;\;\;\;\;\;+ax(27-118B_{1}x+6ax(1-5B_{1})))\Big)\Big],\label{27}
\end{eqnarray}
\begin{eqnarray}
	p_t=\frac{1}{30(1+ax)^5}\Big[-20B_{1}+a^3x^3(81-283B_{1}x)+a^4x^3(51-146B_{1}x\nonumber \\
	\;\;\;\;\;\;+a(30-130B_{1}x)+3a^2x(20-91B_{1}x)-6a^5x^4(5B_{1}-2))\Big],\label{28}
\end{eqnarray}
and
\begin{equation}
	\Delta=p_t-p_r.\label{29}
\end{equation}
\subsection{Regularity at boundary}\label{s3.1}
To determine the unknown quantities required for constructing a physically viable and reliable stellar model, some conditions must be satisfied throughout the star:
\begin{enumerate} 
	\item At the boundary of the star ($r=R$), where $R=$ radius of the star, the interior solution should match with the exterior solution of Schwarzschild given below:
	\begin{eqnarray}
		ds^2&=&-\left(1-\frac{2M}{r}\right)dt^2+\left(1-\frac{2M}{r}\right)^{-1}dr^2  \nonumber \\
		&&  +r^2\left(d{\theta^2}+\sin^2{\theta}d{\phi^2}\right),\label{30}
	\end{eqnarray}
	where $M$ represents the total mass of the star. Matching the interior and exterior metrics at $r=R$, we obtain $e^{\nu(r=R)}=e^{-\lambda(r=R)}=(1-\frac{2M}{R})$.  
	
	\item All the physical parameters associated with the configuration of stellar interior as well as metric potentials should be well behaved. 
	
	\item At the surface of the star the radial pressure should vanish, i.e., $p_r(R )=0$ \cite{mafa}. 
	
	\item The physical parameters such as $\rho$, $p_r$ and $p_t$ should be continuous, positive and monotonically decreasing from the centre to the surface. Moreover, the pressure to density ratio must follow the Zeldovich condition \cite{zeldo, gedala} i.e., $\frac{dp}{d\rho}<1$
	
	\item To maintain overall charge neutrality of the quark matter composition which includes strange quark of finite mass, presence of electrons, though small, are necessary. This puts a restriction on the chemical potential of electron as $\mu_e\geq 0$ for $m_s\neq0$. 
\end{enumerate}

\section{Restriction on coupling constant ($\alpha_c$)}\label{s4}
\begin{figure}[ht]
	\begin{center}
		\includegraphics[width=11cm]{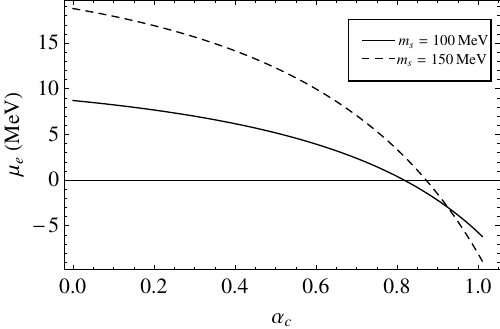}
		\caption{Variation of chemical potential of electron ($\mu_e$) inside the star with coupling parameter ($\alpha_c$) for two different values of $m_s$ taking $B_g=57.55~MeV/fm^3$. Here the solid and dashed lines represent $m_s=100$ and $150~MeV$, respectively.}\label{fig4}
	\end{center}
\end{figure}

In this work, we consider a spherically symmetric static stellar model composed mainly of deconfined $3$-flavour quarks ($u$, $d$ and $s$) and is assumed to be maintained charge neutrality condition in presence of electrons. From \fref{fig3}, $\mu_e$ decreases with increasing $\alpha_c$ for a given $m_s$. Therefore, it may be possible that for a particular value of $\alpha_c$, $\mu_e$ becomes negative. Since $\mu_e$ cannot be negative, some upper limit of the coupling parameter $\alpha_c$ may exist depending on $m_s$ and $B_g$. This particular value of $\alpha_c$ is considered to be the maximum allowed value $(\alpha_{c,max})$. The variation of $\mu_e$ with $\alpha_c$ is shown in \fref{fig4} for two values of $m_s=100$ and $150~MeV$ and $B_g=57.55~MeV/fm^3$. We also observe that with increasing $m_s$, $\alpha_{c,max}$ increases.
\begin{figure}[ht]
	\begin{center}
		\includegraphics[width=11cm]{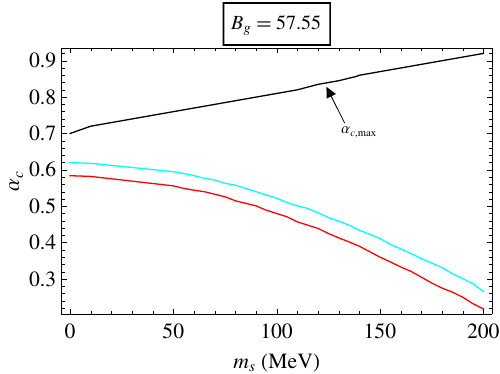}
		\caption{Plot of $\alpha_c$ with $m_s$ for $B_g=57.55~MeV/fm^3$. The black line indicates the maximum allowed value of $\alpha_c$ when $m_s$ increases. The region below the line in red indicates the window for the stable region, the strip between the red and cyan colours represents metastable region, and the portion between the cyan and black lines represents the unstable region.}\label{fig5}
	\end{center}
\end{figure}
Thus, depending on the values of $B_g$ and $m_s$, $\alpha_{c,max}$ has some upper limit, below which the model is physically allowed to study the nature of the SQM inside the star since for such combination $\mu_e\geq0$. When $m_s=100~MeV$, $\alpha_{c,max}=0.81$ and for $m_s=150~MeV$, $\alpha_{c,max}=0.86$. In \fref{fig5}, the black line represents variation of $\alpha_{c,max}$ with $m_s$. Although all values within this line are allowed, only certain values will be physically acceptable. The region below the red line in \fref{fig5} indicates the window for stable SQM ($E_B<930.4~MeV$). The region between the red and cyan lines indicates the metastable window, and the region between cyan and black lines is termed the unstable window, with $E_B>939~MeV$. Although $\alpha_{c,max}$ increases with increasing $m_s$, the area of the stable window decreases. Therefore, thus far, the stability of the SQM has been considered, and to maintain charge neutrality, the values of both $m_s$ and $\alpha_c$ should be small. It is interesting to note that within the parameter space discussed above, a star may be a pure strange quark star composed of stable SQM ($E_B\leq930.4~MeV$) or a hybrid type composed of a mixed phase (hadrons and quarks) of matter for which $E_B$ lies in the range $930.4~MeV<E_B<939~MeV$ or a pure hadronic or neutron star with $E_B>939~MeV$.
\section{Calculation of the maximum mass and surface redshift}\label{s5}
\begin{figure}[ht]
	\begin{center}
		\includegraphics{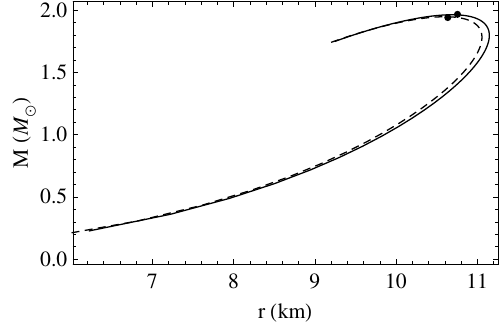}
		\caption{Variation of mass with radius for two independent values of $\alpha_c$, $m_s=100~MeV$ and $B_g=57.55~MeV/fm^3$. The solid and dashed lines represent $\alpha_c=0.0$ and $0.6$, respectively.}\label{fig6}
	\end{center}
\end{figure}

\begin{figure}[ht]
	\begin{center}
		\includegraphics{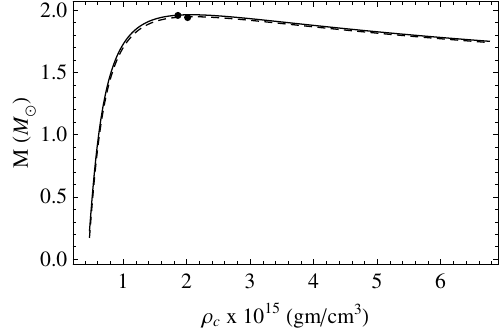}
		\caption{Variation of mass with central density for two values of $\alpha_c$, choosing $m_s=100~MeV$ and $B_g=~57.55~MeV/fm^3$. The solid and dashed lines represent $\alpha_c=0.0$ and $0.6$, respectively.}\label{fig7}
	\end{center}
\end{figure}

\begin{figure}[ht]
	\begin{center}
		\includegraphics{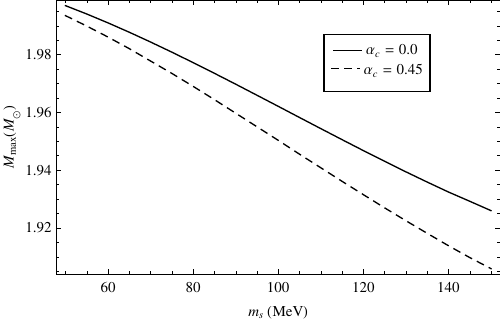}
		\caption{Variation of maximum mass ($M_{max}$) with $m_s$ for $B_g=57.55~MeV/fm^3$. Here, the solid and dashed lines represent values of $\alpha_c=0.0$ and $0.45$, respectively.}\label{fig8}
	\end{center}
\end{figure}

The total mass ($M$) contained within the spherically symmetric and static sphere of radius $R$ in this model can be determined by matching the interior and exterior solutions at the boundary as follows:
\begin{equation}
	M=\frac{R}{2}(1-e^{-\lambda(R)}).\label{31}
\end{equation} 
From equation \eref{31}, it is evident that the total mass depends on the strange quark mass ($m_s$), the coupling parameter ($\alpha_c$) and $B_g$ through the equations \eref{19} and \eref{25}. To study the mass-radius relation of compact objects, we solve the TOV equation considering the EoS of massive and interacting quarks, as given in equation \eref{12}. To estimate the maximum mass we use different parametric combinations of our model parameters, and all the results are tabulated in \tref{tab2}. It is evident that the maximum mass and radius depend on $\alpha_c$, $B_g$ and $m_s$. The $M-R$ variation is shown in \fref{fig6}. The black dots indicate the maximum mass point ($M_{max}$) on each curve. In \fref{fig7}, the variation of mass ($M$) with central density ($\rho_c$) is shown, and it is observed that Harrison-Zeldovich-Novikov stability criterion \cite{BKH,YBZ} holds up to a point for which $\frac{\partial{M}}{\partial{\rho_c}}>0$, referred to as the maximum mass point. In \fref{fig8}, we plotted the maximum mass as a function of $m_s$ for two different parametric choices of $\alpha_c$. \Fref{fig8} shows that the maximum mass decreases with increasing $m_s$ and that the slope is steeper for higher $\alpha_c$ values. 

\begin{table}[t]
	\centering
	\caption{\label{tab2}Table for maximum mass ($M_{max}$), radius ($R_{max}$) and central density ($\rho_{c,max}$), compactness ($u_{max}$) and surface redshift ($Z_{s,max}$) for different choices of $m_s$, $\alpha_c$ and $B_g$. Here, the value of the coupling parameter $\alpha_c$ in bold font corresponds to the maximum possible value ($\alpha_{c,max}$) for a given $m_s$ and $B_g$.}
	\resizebox{0.6\textheight}{!}{$
		\begin{tabular}{@{}c|c|cccccc}
			\hline
			$B_g$                       & $m_s$                &$\alpha_c$       &$M_{max}$    & $R_{max}$    & $\rho_{c,max}~\times{10^{15}}$ & $u_{max}$             & $Z_{s,max}$\\
			($MeV/fm^3$)                & ($MeV$)               &                 &($M_{\odot}$)& ($km$)       & ($gm/cc$)         & ($\frac{M_{max}}{R_{max}}$)          &  \\ \hline
			\multirow{6}{*}{57.55}     &  \multirow{3}{*}{50}      & 0.0             &2.001         & 10.89       &1.99             & 0.2710                 & 0.4777         \\
			                           &                           & 0.6             &1.992         & 10.85       &2.01             & 0.2708                 & 0.4770 \\
			                           &                           & \bfseries{0.76} &1.991         &10.84        &2.01            & 0.2709                  & 0.4774 \\  \cline{2-8}
			                           &\multirow{3}{*}{100}       & 0.0             &1.962         &10.70        &2.05             & 0.2705                 & 0.4759\\
			                           &                           & 0.5             &1.951         &10.62        &2.09             & 0.2710                 & 0.4774\\
			                           &                           & \bfseries{0.81} &1.940         &10.56        &2.091            & 0.2709                 & 0.4776\\ \hline
			\multirow{6}{*}{60}        &\multirow{3}{*}{50}        & 0.0             & 1.956        &10.65        &2.09           &0.2707 &0.4768 \\
			                           &                           & 0.45            &1.952         & 10.64       &2.091           &0.2706 &0.4763  \\
		   	                           &                           & \bfseries{0.76} &1.950         &10.63        &2.092            &0.2705 &0.4762 \\ \cline{2-8}
			                           &\multirow{3}{*}{100}       & 0.0             &1.922         &10.48        &2.153            & 0.2705                 & 0.4761\\
			                           &                           & 0.45            &1.911         &10.43        &2.152            & 0.2703                 & 0.4752 \\
			                           &                           & \bfseries{0.8}  &1.902         &10.38        &2.172            & 0.2702                 & 0.4753 \\ \hline                       
		\end{tabular}$}
\end{table}

The total mass ($M$), radius ($R$), baryon number density ($n_i$) and the moment of inertia ($I$) for slow and rigid rotation are some the important parameters of a stellar  system from a statistical point of view. In addition, there is another parameter called "surface redshift" ($Z_s$), which accounts for the change in frequency of a photon travelling radially from the stellar surface to infinity. The expression for the surface redshift $Z_s$ is \cite{GHB}

\begin{equation}
	Z_s=\left(1-\frac{2M}{R}\right)^{-1/2}-1\label{32}
\end{equation}
For maximum mass and radius and following equation \eref{32}, it is noted that the surface redshift is also maximum. . 

\section{Physical application}\label{s6}
For physical application, we consider compact objects with known masses ($M$) and study the effect of the coupling parameter ($\alpha_c$) in the presence of nonzero $m_s$ on various physical properties of the star in the stable region. Alhough for stable strange matter relative to neutrons, the range of $B_g$ is $57.55~MeV/fm^3<B_g<95.11~MeV/fm^3$, it is modified in the presence of a finite mass of strange quarks ($m_s\neq0$) and a coupling parameter ($\alpha_c$), as given in \tref{tab1}. We focus mainly on the stable region for physical interpretation. Since for each $m_s$ there is an upper bound of $\alpha_c$, we can tune $m_s$, $\alpha_c$ and $B_g$ within this limit, to estimate various physical properties such as the radius, central pressure, central density, etc. To analyse our model, we choose two compact stars, namely, $(1)~PSR~J1614-2230$, whose estimated mass and radius from observations are $1.97~M_{\odot}$ and $9.69~km$ \cite{TG}, respectively and $(2)~EXO~1745-268$ with estimated mass $1.65~M_{\odot}$ and radius of $10.5~km$ \cite{FO}. Now, we discuss the two cases separately.\par
\begin{figure}[ht]
	\begin{center}
		\includegraphics{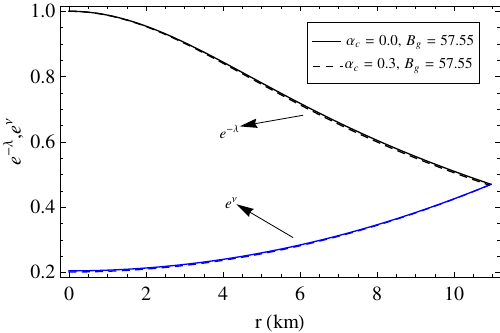}
		\caption{Radial variation of $e^{-\lambda}$ and $e^{\nu}$, for $m_s=~100~MeV$ and $B_g=57.55~MeV/fm^3$ in $PSR~J1614-2230$. The solid and dashed lines indicate $\alpha_c=0.0$ and $0.3$, respectively.}\label{fig9}
	\end{center}
\end{figure}

\begin{figure}[ht]
	\begin{center}
		\includegraphics{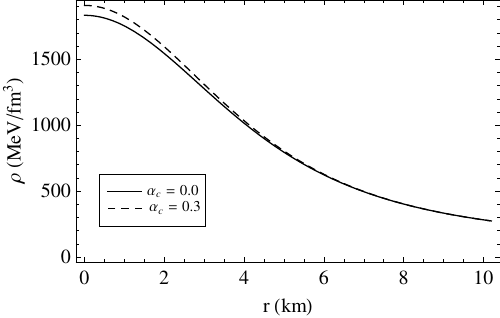}
		\caption{Variation of $\rho$ with $r$ inside $PSR~J1614-223$ for different $\alpha_c$, taking a parametric choice of $m_s=100~MeV$ and $B_g=65~MeV/fm^3$. Here the solid and dashed lines represent values of $\alpha_c=0.0$ and $0.3$, respectively.}\label{fig10}
	\end{center}
\end{figure}

\begin{figure}[ht]
	\begin{center}
		\includegraphics{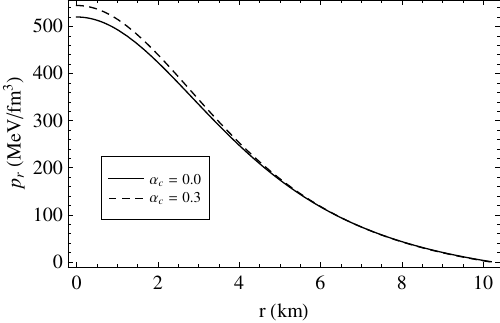}
		\caption{Radial variation of $p_r$ inside $PSR~J1614-2230$ for $\alpha_c=0.0$ and $0.3$, represented respectively by the solid and dashed lines with a parametric choice of $m_s=100~MeV$ and $B_g=65~MeV/fm^3$}\label{fig11}
	\end{center}
\end{figure}

\begin{figure}[ht]
	\begin{center}
		\includegraphics{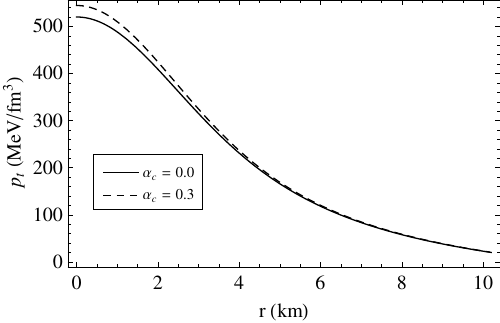}
		\caption{Variation of $p_t$ with $r$ inside $PSR~J1614-2230$ for different $\alpha_c$, taking a parametric choice of $m_s=100~MeV$ and $B_g=65~MeV/fm^3$. Here the solid and dashed lines represent the choices $\alpha_c=0.0$ and $0.3$, respectively.}\label{fig12}
	\end{center}
\end{figure}

\begin{table}
	\centering
	\caption{\label{tab3}Predicted radii of different compact objects with known masses.}
	\resizebox{0.8\textwidth}{!}{$
		\begin{tabular}{@{}ccc|cccc}
			\hline
			Name of the & Observed      & Estimated        & \multicolumn{4}{c}{Predicted radius}  \\ \cline{4-7}
			stars       &   Mass        &  radius          & $\alpha_c$  & $B_g$  & $m_s$   & R \\
			&  ($M_{\odot}$)&     ($km$)       &             & ($MeV/fm^3$) & ($MeV$) &  ($km$) \\ \hline
			$PSR~J1614-2230$ & 1.97\cite{TG} & 9.69$\pm$0.2 & 0.23 & 69.5 & 100 & 9.67\\
			$4U~1608-52$     & 1.57\cite{FO}  & 9.8$\pm$1.8  & 0.25  & 71 & 108 & 9.81 \\
			$Vela~X-1$       & 1.77\cite{TG}  & 9.56$\pm$0.08  & 0.18 & 77 & 70 & 9.55 \\
			$EXO~1745-268$   & 1.65\cite{FO}  & 10.5$\pm$1.6 & 0.4 & 63 & 65 & 10.55 \\
			$PSR~J1903+327$  & 1.667\cite{TG} & 9.438$\pm$0.03 & 0.14 & 79 & 70 & 9.48\\ \hline            
		\end{tabular}$}
\end{table}
\begin{table}
	\centering
	\caption{\label{tab4} Values of central pressure ($p_{r0}$), central density ($\rho_c$), surface density ($\rho_s$) and nature of SQM}
	\resizebox{0.6\textwidth}{!}{$
		\begin{tabular}{@{}c|ccccc}
				\hline
			Name of the                       & $\alpha_c$ & $p_{r0}\times{10^{35}}$   & $\rho_c \times{10^{15}}$  & $\rho_s \times{10^{14}}$  & Nature of \\
			compact object                    &            & ($dyne/cm^2$)           & ($gm/cc$)               & ($gm/cc$)               &  SQM  \\ \hline
			\multirow{2}{*}{$PSR~J1614-2230$} & 0.0        & 14.2                    &  5.24                   & 5.20                    & stable\\
			                                  & 0.23       & 15.0                    & 5.51                    &5.25                     & stable\\  \cline{2-6}
			\multirow{2}{*}{$4U~1608-52$}     & 0.0        & 3.12                    & 1.57                    & 5.32                    & stable\\
			                                  & 0.25       & 3.16                    & 1.59                    & 5.35                    & metastable\\ \cline{2-6}
			 \multirow{2}{*}{$Vela~X-1$}      & 0.0        & 6.26                    & 2.63                    & 5.48                    & stable \\
			                                  & 0.18       & 7.12                    & 2.94                    & 5.62                    & stable \\ \cline{2-6}
			\multirow{2}{*}{$EXO~1745-268$}   & 0.0        & 2.51                    & 1.30                    & 4.59                    & stable \\
			                                  & 0.4        & 2.54                    & 1.31                    & 4.61                    & stable \\ \cline{2-6}
			 \multirow{2}{*}{$PSR~J1903+327$} & 0.0        & 5.33                    & 2.35                    & 5.76                    & stable \\ 
			                                  & 0.14       & 5.36                    & 2.36                    & 5.76                    & stable \\ \hline
		\end{tabular}$}
\end{table}

\begin{figure}[ht]
	\begin{center}
		\includegraphics{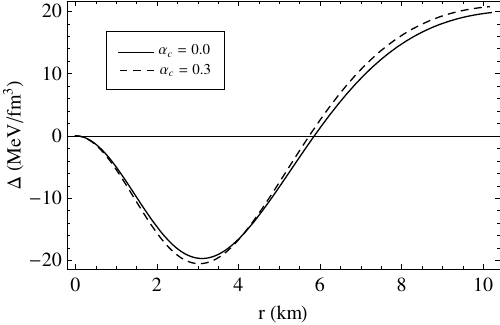}
		\caption{Radial variation of $\Delta$ inside $PSR~J1614-2230$ for different $\alpha_c$, taking a parametric choice of $m_s=100~MeV$ and $B_g=65~MeV/fm^3$. Here the solid and dashed lines represent values of $\alpha_c=0.0$ as and $0.3$, respectively.}\label{fig13}
	\end{center}
\end{figure}

\textbf{Case 1:} We investigate the effects of $B_g$, $m_s$ and $\alpha_c$ on the massive pulser $PSR~J1614-2230$. In \fref{fig9}, the variations of metric potentials ($e^{-\lambda}$, $e^{\nu}$) are shown for different values of $\alpha_c$, where $m_s=100~MeV$ and $B_g=57.55~MeV/fm^3$. Notably $e^{-\lambda}$ at centre and $e^{\nu}$ at the surface are independent of $\alpha_c$. Although $e^{-\lambda}$ at the surface changes with $\alpha_c$. All the relevant physical parameters, such as, energy density ($\rho$), radial and transverse pressures ($p_r, p_t$), and anisotropy ($\Delta$), are evaluated graphically for different parametric combinations of $m_s$, $\alpha_c$ and $B_g$, and the variations are shown in \fref{fig10} -- \fref{fig13}, respectively. It is evident from \fref{fig10} --  \fref{fig12} that $\rho$, $p_r$ and $p_t$ are positive throughout the interior, maximum at the centre and decrease gradually towards the surface. From \fref{fig10} -- \fref{fig12}, we see that the density, and the radial and transverse pressures increase with increasing $\alpha_c$ for a given value of $m_s$ and $B_g$. \Fref{fig13} shows that $\Delta$ is zero at the centre, increases in the negative direction up to a certain maximum and reverses its sign at a particular radial distance. This type of variation in the anisotropic profile has also been previously mentioned by some authors \cite{Zubair,PBHAR}. 

 \begin{figure}[ht]
	\begin{center}
		\includegraphics{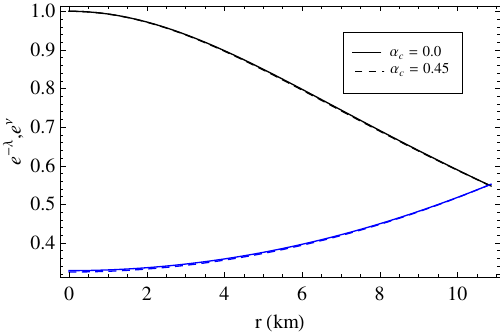}
		\caption{Radial variation of $e^{-\lambda}$ and $e^{\nu}$, for $m_s=100~MeV$ and $B_g=57.55~MeV/fm^3$ in $EXO~1745-268$. The solid and dashed lines indicate $\alpha_c=0.0$ and $0.45$, respectively.}\label{fig14}
	\end{center}
\end{figure}

\begin{figure}[ht]
 	\begin{center}
 		\includegraphics{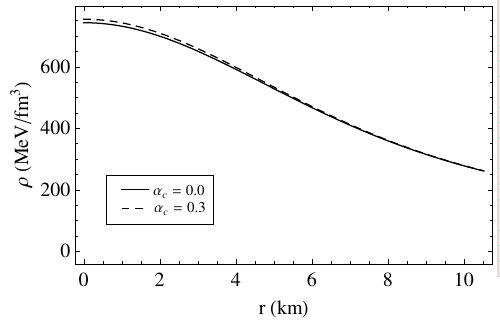}
 		\caption{Radial variation of $\rho$ inside $EXO~1745-268$ for different $\alpha_c$, taking a parametric choice of $m_s=100~MeV$ and $B_g=62~MeV/fm^3$.}\label{fig15}
 	\end{center}
 \end{figure}

\begin{figure}[ht]
 	\begin{center}
 		\includegraphics{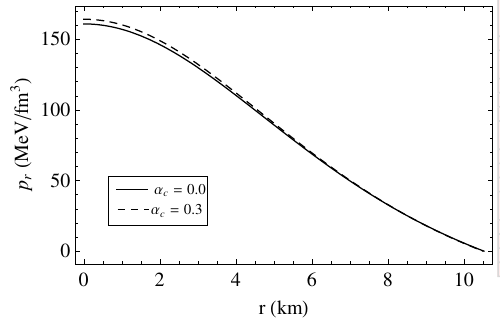}
 		\caption{Radial variation of $p_r$ inside $EXO~1745-268$ for different $\alpha_c$, with a parametric choice of $m_s=100~MeV$ and $B_g=62~MeV/fm^3$.}\label{fig16}
 	\end{center}
 \end{figure}

\begin{figure}[ht]
	\begin{center}
		\includegraphics{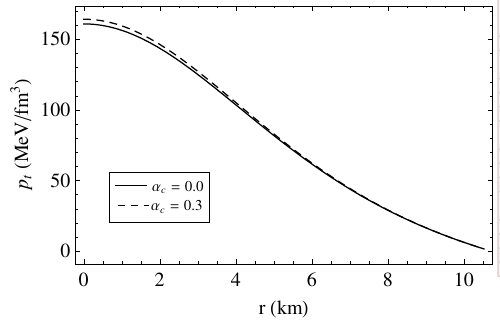}
		\caption{Radial variation of $p_t$ inside $EXO~1745-268$ for different $\alpha_c$, for a parametric choice of $m_s=100~MeV$ and $B_g=62~MeV/fm^3$}\label{fig17}
	\end{center}
\end{figure}

\begin{figure}[ht]
	\begin{center}
		\includegraphics{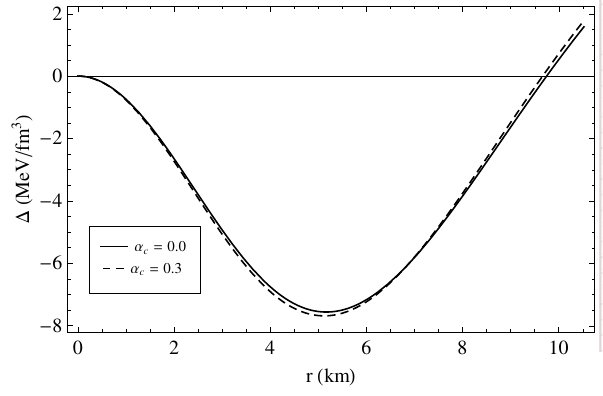}
		\caption{Radial variation of $\Delta$ inside $EXO~1745-268$ for different $\alpha_c$, choosing parametric values of $m_s=100~MeV$ and $B_g=62~MeV/fm^3$}\label{fig18}
	\end{center}
\end{figure}

\textbf{Case 2:} For the next compact star candidate $EXO~1745-268$, with a mass $1.65~M_{\odot}$ and an estimated radius of $10.5~km$ \cite{FO}, we have studied the effects of $\alpha_c$, $m_s$ and $B_g$ on various physical parameters graphically. In \fref{fig14}, the variations of metric potentials ($e^{-\lambda}, e^{\nu}$) are plotted against $r$ for different $\alpha_c$ values with $m_s=100~MeV$ and $B_g=57.55~MeV/fm^3$. In \fref{fig15}--\fref{fig17}, the radial variations of $\rho$, $p_r$ and $p_t$ are shown for the parametric choice of $\alpha_c$, taking $m_s=100~MeV$ and $B_g=62~MeV/fm^3$. These plots indicate that with increasing $\alpha_c$ density, the radial and transverse pressure increase. The radial variation of the anisotropy $\Delta$ is shown in \fref{fig18} which is similar in nature as in case- $1$. 
 
\begin{figure}[ht]
	\begin{center}
		\includegraphics[width=11cm]{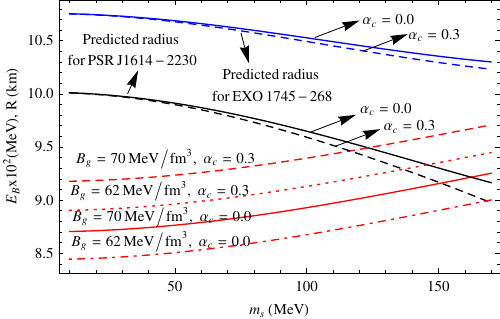}
		\caption{Variation of predicted radius ($R$) and $E_B$ with $m_s$ for different choices of $\alpha_c$. The lines in black and blue represent variation of $R$ for $PSR~J614-2230$ and $EXO~1745-268$, taking different parametric choices of $\alpha_c$, $B_g=70~MeV/fm^3$ and $62~MeV/fm^3$, respectively. The solid, dashed, dot-dashed and dotted lines in red represent the variation in $E_B$ for the chosen parametric values of $\alpha_c$ and $B_g$.}\label{fig19}
	\end{center}
\end{figure}

\begin{table}[t]
	\centering
	\caption{\label{tab5} Table for compactness ($u=\frac{M}{R}$) and surface redshift ($Z_s$) for the different stars for the predicted radius using the values of \tref{tab3}.}
	\resizebox{0.6\textheight}{!}{$
		\begin{tabular}{@{}cc|ccc}
			\hline
			Name of the & Observed mass & Predicted radius & u   & $Z_s$ \\ 
			compact star & ($M_{\odot}$)& ($km$)           & ($\frac{M}{R}$)     &       \\ \hline
			$PSR~J1614-2230$ & 1.97 & 9.67 & 0.3005 & 0.5831 \\
			$4U~1608-52$  & 1.57 & 9.81 & 0.2361 & 0.3764 \\
			$Vela~X-1$   & 1.77 & 9.70  & 0.2691 & 0.4717 \\
			$EXO~1745-268$ & 1.65 & 10.55 & 0.2307 & 0.3626\\
			$PSR~J1903+327$  & 1.667 & 9.48 & 0.2593 & 0.4414\\ \hline
        \end{tabular}$}
\end{table}
Since $\alpha_c$ has a nominal effect on the mass and radius of a star as evident from \tref{tab2}, and has a major effect on $E_B$, as observed from \fref{fig2}, we have shown in \fref{fig19} the variation of the predicted radius ($R$) and $E_B$ with $m_s$ taking different parametric combinations of $\alpha_c$ and $B_g$ for the two chosen compact stars. For $PSR~J1614-2230$, we choose $B_g=70~MeV/fm^3$, $\alpha_c=0.0$, $0.3$ and for $EXO~1745-268$, $B_g=62~MeV/fm^3$ with $\alpha_c=0.0$ and $0.3$ is chosen. It is evident that with increasing $m_s$, the predicted radius decreases, whereas $E_B$ increases, shifting the stellar matter towards the unstable region for both stars. Additionally, this variation is more prominent with higher $\alpha_c$. Therefore, the requirement for lower $m_s$ and $\alpha_c$ is justified for modeling the star in the stable region. In \tref{tab3}, we have tabulated the predicted radius of several compact stars for the suitable choices of$\alpha_c$, $m_s$ and $B_g$. Notably, the predicted radii are almost comparable to the estimated values from the observations. In \tref{tab4}, the nature of the SQM inside the known compact stars, which are assumed to be the strange stars are tabulated, and corresponding values of the central density, pressure and surface density are also shown. \Tref{tab5} show the values of $Z_s$ and compactness, i.e., the mass to radius ratio ($u=\frac{M}{R}$) of the compact objects for the combinations of parameters used in \tref{tab3}. 

\subsection{Causality condition}\label{s6.1}
\begin{figure}[!htb]
	\begin{minipage}{0.55\textwidth}
		\centering
		\includegraphics[width=0.95\linewidth,height=0.8\textwidth]{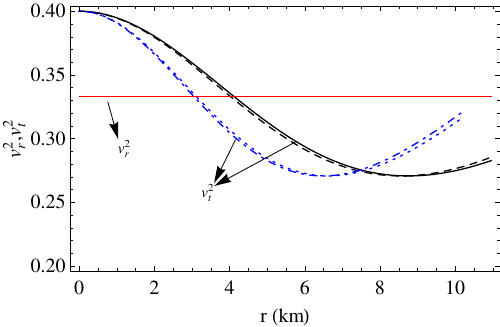}
		\caption{Variation of $v_r^2$ and $v_t^2$ sound velocities with $r$ in $PSR~J1614-2230$ for different combinations of $\alpha_c$, $m_s$ and $B_g$. Here, the red line represents $v_r^2$. The solid and dashed black lines represent the variation of $v_t^2$ for $\alpha_c=0.0$ and $0.45$ with $B_g=57.55~MeV/fm^3$, whereas, the dotted and dotdashed blue lines represent the variation for $\alpha_c=0.0$ and $0.30$ with $B_g=65~MeV/fm^3$.}\label{fig20}
\end{minipage}\hfil
\begin{minipage}{0.55\textwidth}
\centering
		\includegraphics[width=0.95\linewidth,height=0.8\textwidth]{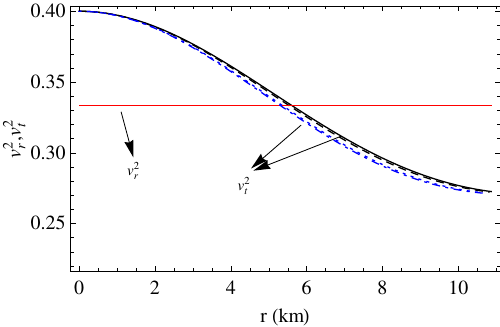}
		\caption{Variation of $v_r^2$ and $v_t^2$ sound velocities with $r$ in $EXO~1745-268$ for different combinations of $\alpha_c$, $m_s$ and $B_g$. Here, the red line represents $v_r^2$. The solid and dashed black lines represent the variation of $v_t^2$ for $\alpha_c=0.0$ and $0.45$ with $B_g=57.55~MeV/fm^3$, whereas, the dotted and dotdashed blue lines represent the variation for $\alpha_c=0.0$ and $0.35$ with $B_g=60~MeV/fm^3$.}\label{fig21}
	\end{minipage}
\end{figure}

To construct a physically viable stellar model, in presence of anisotropy the causality conditions must be satisfied at all internal points and on the boundary. Mathematically this can be expressed as $0<v_r^2\leq1$ and $0<v_t^2\leq1$, where $v_r^2$ and $v_t^2$ are the squares of the radial and transverse sound velocities, respectively. We have shown graphically in \fref{fig20} and \fref{fig21} the values of both the sound velocities inside the compact stars, $PSR~J1614-2230$ and $EXO~1745-268$, for different parametric combinations of $\alpha_c$, $m_s$ and $B_g$, respectively. It is evident that both the sound velocity are within the allowed range. Moreover, $v_r^2$ is always $\frac{1}{3}$, independent of $\alpha_c$, $m_s$ and $B_g$.

\subsection{Energy conditions}\label{s6.2}
Since the exact internal composition of a compact star is still under debate, useful information may be obtained by imposing some energy conditions on the system. These conditions are also helpful in determining a viable energy momentum tensor. In case of astrophysical context, the study of energy conditions in principle is an algebraic problem \cite{CAK}, more specifically it is an eigen value problem which is related to the energy momentum tensor. In particular, if fluids withstand important energy conditions namely, Strong Energy Condition (SEC), Weak Energy condition (WEC), Null Energy Condition (NEC) and Dominant Energy Condition (DEC) then general relativity can not be taken into account. The study of the energy conditions in case of fluid in $4$-dimensional space-time corresponds to the solutions of polynomial of degree $4$ to evaluate the roots. However, it is very complicated because of the difficulty in evaluating the analytical solutions for such eigenvalues. In spite of the difficulty in obtaining the general solution for the roots, the energy conditions \cite{CAK,SWH,RW} should be followed simultaneously by the physically realistic fluid distribution within the stellar configuration. For any fluid distribution that is isotropic or anisotropic in nature, the stated energy conditions must be satisfied. Such energy conditions can be expressed mathematically as \cite{BP,BPB}:

\begin{enumerate}
	\item DEC: $\rho\geq 0,~\rho - p_{r}\geq 0,~\rho - p_{t}\geq 0$.
	\item WEC: $\rho + p_{r}\geq 0,~\rho \geq 0, \rho + p_{t}\geq 0$.
	\item NEC: $\rho + p_{r}\geq 0,~\rho + p_{t} \geq 0$.
	\item SEC: $\rho + p_{r}\geq 0,~\rho + p_{r} + 2p_{t}\geq 0$. 
\end{enumerate}
 \begin{figure}[ht]
 	\centering
 	\includegraphics[width=10cm]{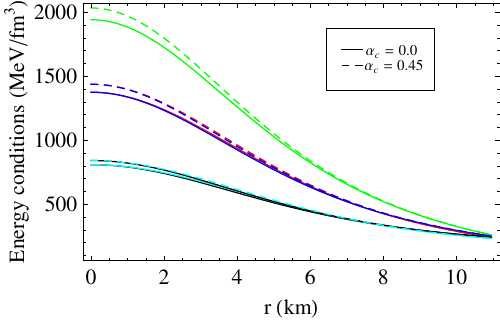}
 	\caption{Variation of different energy conditions inside $PSR~J1614-2230$ with $r$ for different parametric choices of $\alpha_c$. The red, blue, green, black and cyan lines indicate $(\rho+p_r)$, $(\rho+p_t)$, $(\rho+p_r+2p_t)$, $(\rho-p_r)$ and $(\rho-p_t)$, respectively.}\label{fig22}
 \end{figure}
  \begin{figure}[ht]
 	\centering
 	\includegraphics[width=10cm]{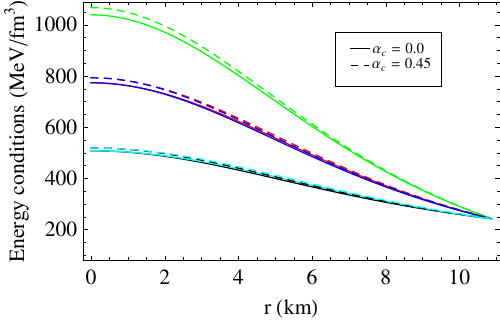}
 	\caption{Variation of different energy conditions inside $EXO~1745-268$ with $r$ for different parametric choices of $\alpha_c$. The red, blue, green, black and cyan lines indicate $(\rho+p_r)$, $(\rho+p_t)$, $(\rho+p_r+2p_t)$, $(\rho-p_r)$ and $(\rho-p_t)$, respectively.}\label{fig23}
 \end{figure}
In \fref{fig22} and \fref{fig23}, the different energy conditions are shown for the selected compact objects with different combinations of model parameters. These plots clearly show that all the necessary energy conditions are satisfied in the interior of the stellar configuration. 

\subsection{Moment of inertia ($I$)}\label{s6.3}
In the modeling of compact stars such as pulsars, the moment of inertia ($I$) plays an important role. The relation between moment of inertia $I$ and matter distribution is very complex due to the fact of gneral relativistic effects such as the dragging of local inertial frame. Additionally, $I$ depends sensitively on the EoS. To derive the relation between mass and moment of inertia ($M-I$) in this model, an approximate expression of $I$ of a star is used, as proposed by Bejger and Haensel \cite{BH}, which shows that a slowly rotating solution can be transformed into a static one via the expression of $I$ given below:
  \begin{equation}
  	I=\frac{2}{5}\left(1+\frac{(M/R).km}{M_\odot}\right)MR^2.\label{33}
  \end{equation}
\begin{figure}[ht]
	\begin{center}
		\includegraphics{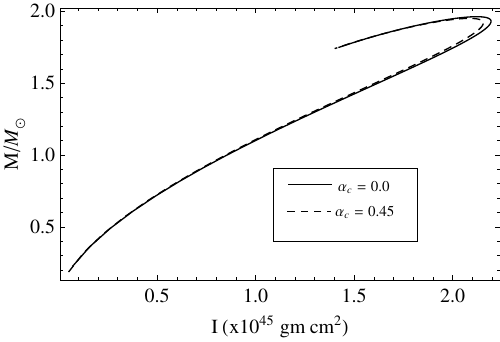}
		\caption{Variation of $I$ with mass $M$ for $m_s=100~MeV$ and $B_g=57.55~MeV/fm^3$. Here the solid and dashed lines indicate $\alpha_c=0.0$ and $0.45$, respectively.}\label{fig24}
	\end{center}
\end{figure}
Using equation \eref{33} and with a parametric choice of $m_s$, $B_g$ we graphically show the variation of the moment of inertia ($I$) for different $\alpha_c$ values in \fref{fig24}. The calculated values of $I$ for the mass and predicted radius of compact stars of known mass and radius are shown in \tref{tab6}. 

\begin{table}
	\centering
	\caption{\label{tab6}Approximate values of the moment of inertia ($I$) for different compact objects with parametric choices of $\alpha_c$, $m_s$ and $B_g$.}
	\resizebox{0.6\textheight}{!}{$
	    \begin{tabular}{@{}cccc}
			\hline
		Name of the      & Observed mass & Predicted radius  &  moment of inertia ($I$) \\ 
		compact stars    & ($M_\odot$)    &  ($km$)  & $\times10^{45}~(gm.cm^2)$        \\ \hline 
		$PSR~J1614-2230$ & 1.97     &  9.67  & 1.774                                    \\
	    $4U~1608-52$     & 1.57     &  9.81  & 1.402                                    \\ 
	    $Vela~X-1$       & 1.77     &  9.70  & 1.575                                 \\   
	    $EXO~1745-268$   & 1.65     &  10.55  & 1.699                                   \\ 
	    $PSR~J1903+327$  & 1.667    &  9.48    & 1.409                                  \\ \hline
			\end{tabular}$} 
	\end{table}

\section{Stability analysis}\label{s7}
The following stability analysis are performed to check the physical viability of present model
\begin{enumerate}
	\item Study of stability under the TOV equation
	\item Cracking condition proposed by Herrera 
	\item Evaluation of the adiabatic index and
	\item Lagrangian change in radial pressure with respect to small radial oscillation
\end{enumerate}

\subsection{Study of stability under the TOV equation}\label{s7.1}
The generalised TOV equation \cite{tolman,Oppenheimer} is represented as 
\begin{equation}
	-\frac{M_{G}(r)(\rho+p_{r})}{r^2}e^{(\lambda-\nu)/2}-\frac{dp_{r}}{dr}+\frac{2}{r}(p_{t}-p_{r})=0.\label{34}
\end{equation}
where $e^{\nu}$ and $e^{\lambda}$ are the metric potentials and are given in equations \eref{24} and \eref{25}. $M_{G}(r)$ is the effective gravitational mass contained within a spherical region of radius $r$ and can be obtained from the formula of Tolman-Whittaker and using the EFE as:
\begin{equation}
	M_{G}(r)=\frac{1}{2}r^{2}\nu^{\prime}e^{(\nu-\lambda)/2}.\label{35}
\end{equation}
Substituting equation \eref{35} into equation \eref{34}, the following relation can be derived,
\begin{equation}
	-\frac{\nu^{\prime}}{2}(\rho+p_{r})-\frac{dp_{r}}{dr}+\frac{2}{r}(p_{t}-p_{r})=0.\label{36}
\end{equation}

\begin{figure}[!htb]
	\begin{minipage}{0.55\textwidth}
		\centering
		\includegraphics[width=0.95\linewidth,height=0.8\textwidth]{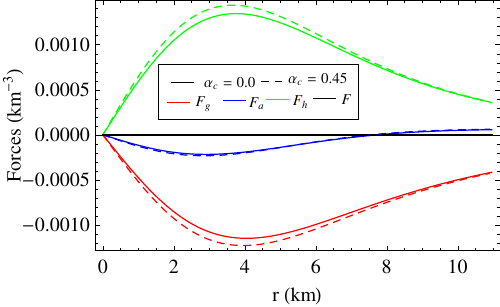}
		\caption{Variation of different forces with $r$ inside $PSR~J1614-2230$ for parametric choices of $m_s=~100~MeV$ and $B_g=57.55~MeV/fm^3$. The solid and dashed lines indicate $\alpha_c=0.0$ and $0.45$, respectively. Here, the red, blue and green lines indicate $F_g$, $F_a$ and $F_h$, respectively.}\label{fig25}
	\end{minipage}\hfil
\begin{minipage}{0.55\textwidth}
\centering
		\includegraphics[width=0.95\linewidth,height=0.8\textwidth]{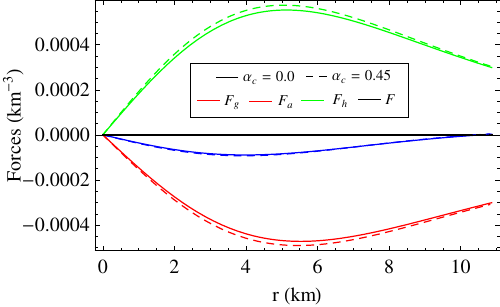}
		\caption{Variation of different forces with $r$ inside $EXO~1745-268$ for parametric choices of $m_s=~100~MeV$ and $B_g=57.55~MeV/fm^3$. The solid and dashed lines indicate $\alpha_c=0.0$ and $0.45$, respectively. Here, the red, blue and green lines indicate $F_g$, $F_a$ and $F_h$, respectively.}\label{fig26}
	\end{minipage}
\end{figure}

The TOV equation as given in equation \eref{36}, represents the equilibrium condition in stellar interior, under the combined effect of forces which are, gravity force ($F_{g}$), the hydrostatic force ($F_{h}$) and the force due to pressure anisotropy ($F_{a}$), given respectively as $F_g=-\frac{\nu^{\prime}}{2}(\rho+p_{r})$, $F_h=-\frac{dp_{r}}{dr}$ and $F_a=\frac{2\Delta}{r}$. Using the expressions of $F_g$, $F_h$ and $F_a$ one can write that the sum of all three forces inside the star is zero always, i.e., $F_g+F_h+F_a=0$. In \fref{fig25} and \fref{fig26}, we have shown the variation of these forces inside $PSR~J1614-2230$ and $EXO~1745-268$, respectively, with $r$ for different parametric choices of model parameters and we note that the static equilibrium condition holds good inside the stars. We also note that the total effect of the hydrostatic force $(F_h)$ is balanced by anisotropic $(F_a)$ and the gravity force ($F_g$) in the presence of nonzero $m_s$ and $\alpha_c$. 

\subsection{Herrera cracking condition}\label{s7.2}
Any compact object of isotropic or anisotropic in nature must be stable with respect to small changes in their physical variables. The idea of cracking was introduced by Herrera \cite{Herrera2} to examine whether an anisotropic fluid configuration for self gravitating objects is stable or not. Based on the concept of Herrera \cite{Herrera2}, a criterion has been proposed by Abreu \cite{Abreu}. According to Abreu \cite{Abreu}, if the square of the radial ($v_{r}^{2}$) and tangential ($v_{t}^{2}$) sound speeds satisfy the following condition inside a star, then it is possible to say that the stellar model is in stable equilibrium.    
\begin{equation}
	0 \leq |v_{t}^{2}-v_{r}^2|\leq {1}. \label{37}
\end{equation}

\begin{figure}[ht]
	\begin{center}
		\includegraphics[width=10cm]{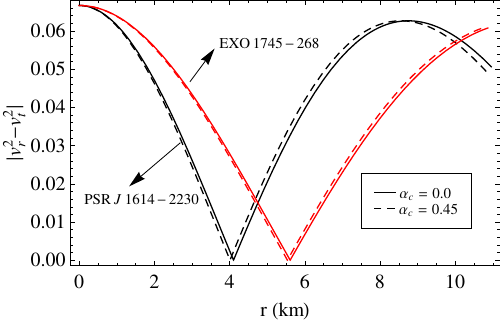}
		\caption{Radial variation of $|v_r^{2}-v_t^{2}|$ inside $PSR~J1614-2230$ and $EXO~1745-268$ for the parametric choices of $m_s=100~MeV$ and $B_g=57.55~MeV/fm^3$, as indicated by the black and red lines, respectively. Here, the solid and dashed lines indicate $\alpha_c=0.0$ and $0.45$, respectively.}\label{fig27}
	\end{center}
\end{figure}

From \fref{fig27}, it is verified that both compact stars remain causal for different parametric choices of $\alpha_c$, $m_s$ and $B_g$.

\subsection{Adiabatic index}\label{7.3}
In case of a star of isotropic fluid the adiabatic index ($\Gamma$) is defined as: 
\begin{equation}
	\Gamma=\frac{\rho+p_{r}}{p_{r}}\frac{dp_{r}}{d\rho}=\frac{\rho+p_{r}}{p_{r}}v_{r}^{2}. \label{38}
\end{equation}
For stable isoptropic fluids, Heintzmann and Hillebrandt \cite{HH} evaluated that $\Gamma>\frac{4}{3}$ (Newtonian limit). However, for a relativistic anisotropic fluid, Chan et al. \cite{CHEN}  has derived a new limit of $\Gamma$  which given by $\Gamma>\Gamma^{\prime}_{max}$, where,
\begin{equation}
	\Gamma^{\prime}_{max}=\frac{4}{3}-\left[\frac{4}{3}\frac{p_r-p_t}{|p_r^{\prime}|r}\right]_{max}. \label{39}
\end{equation}

\begin{figure}[ht]
	\begin{center}
		\includegraphics{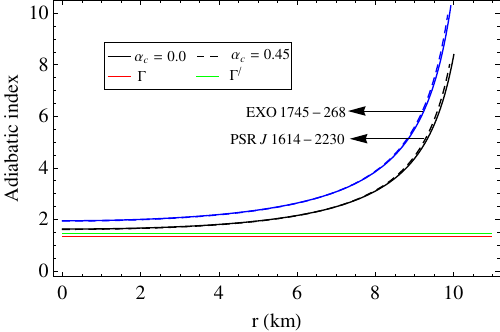}
		\caption{Variation of the adiabatic index with $r$ inside $PSR~J1614-2230$ and $EXO~1745-268$, respectively shown by black and blue lines, with the chosen parameters $m_s=100~MeV$ and $B_g=57.55~MeV/fm^3$. Here, the solid and dashed lines indicate the variations for $\alpha_c=0.0$ and $0.45$, respectively. The lines in green and red  indicate $\Gamma^{\prime}$ and $\Gamma$, respectively.}\label{fig28} 
	\end{center}
\end{figure}

In our model, we have calculated $\Gamma$ from equation \eref{38} using different values of $\alpha_c$ and graphically show the variation for $PSR~J1614-2230$ and $EXO~1745-268$ in \fref{fig28}. The figure shows that $\Gamma$ is always greater than $\Gamma^{\prime}_{max}$, resulting in a stable model.

\subsection{Variation of Lagrangian change in the radial pressure with the frequency of normal modes of oscillation}\label{s7.4}
The stability of a stellar system can also be predicted by the Lagrangian change in the radial pressure at the surface of the star with respect to the frequency ($\omega^2$) of normal modes. Such behaviour cn be studied by introducing small perturbations on the radial pressure and the corresponding frequencies of vibrations ($\omega_{0}^2$) of the normal modes can be evaluated. Assuming that the vibrations are adiabatic, according to \cite{pretel1}, the coupled equations associated with the infinitesimal oscillations of radial mode are given below:
\begin{equation}
	\frac{d\zeta}{dr}=-\frac{1}{r}\left(3\zeta+\frac{{\Delta}p_r}{{\gamma}p_r}\right)+\frac{1}{2}\frac{d{\nu}}{dr}\zeta,\label{40}
\end{equation}
\begin{eqnarray}
	\frac{d({\Delta}p_r)}{dr}&=&\zeta\left[\frac{\omega^2}{c^2}e^{(\lambda-\nu)}(\rho+p_r)r-4\frac{dp_r}{dr}\right. \nonumber \\
	&& \left.-k(\rho+p_r)e^{\lambda}rp_r+\frac{r}{4}(\rho+p_r)\left(\frac{d\nu}{dr}\right)^2\right] \nonumber \\
	&& -{\Delta}p_r\left[\frac{1}{2}\frac{d\nu}{dr}+\frac{k}{2}(\rho+p_r)re^{\lambda}\right],\label{41}
\end{eqnarray}
where $k=\frac{8{\pi}G}{c^4}=1$ and $\left|{\Delta}p_r\right|$ is the absolute value of the Lagrangian perturbation in radial pressure. $\zeta(r)$ represents the eigenfunction which is given by $\zeta(r)=\frac{\xi(r)}{r}$, where $\xi(r)$ is the Lagrangian displacement. However, owing to spherical symmetry, the Lagrangian displacement should vanish at the stellar centre, i.e., $\xi(0)=0$. Following the work of Pretel \cite{pretel1}, we consider the approach where $\zeta(r)$ is normalised, i.e., $\zeta(0)=1$. Furthermore, equation \eref{40} poses a singularity at $r=0$. Now, to solve equations \eref{40} and \eref{41}, one requires that the coefficient of $(\frac{1}{r})$ in equation \eref{40} must vanish as $r\rightarrow0$. This implies,
\begin{equation}
	{\Delta}p_r=-3{\gamma}{\zeta}p_r.\label{42}
\end{equation}
Also, the radial pressure is zero at the surface of the star, implementing the boundary condition that, as $r\rightarrow{R}$, $\left|{\Delta}p_r(R)\right|=0$.

\begin{figure}[ht]
	\begin{center}
		\includegraphics{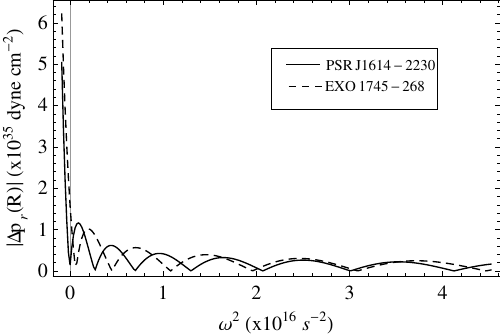}
		\caption{Plots of $\left|p_r(R)\right|$ with $\omega^2$ for $m_s=100~MeV$ and $\alpha_c=0.25$. Here, the solid and dashed lines indicate variation for $PSR~J1614-2230$ and $EXO~1745-268$, respectively for $B_g=65~MeV/fm^3$.}\label{fig29}
	\end{center}
\end{figure}
We have graphically solved the coupled equations \eref{40} and \eref{41} to show the variation of the absolute value of the Lagrangian change in the radial pressure with $\omega^2$ for the selected stars in \fref{fig29} for a parametric choice of $m_s$ and $\alpha_c$. As observed from \fref{fig29} that all the normal mode frequencies correspond to $\omega^2>0$, indicating stability of the model.
 
\section{Discussion}\label{s8}
In this work, we have focused on the effect of the QCD coupling constant ($\alpha_c$ ) on the stellar configuration in case of nonzero value of strange quark mass ($m_s$). The internal matter consisting of $3$-flavour quarks ($u$, $d$ and $s$) and electrons ($e^{-1}$) is overall charge neutral. Considering first order correction in coupling constant, the thermodynamic potentials of the constituent particles are modified and given in equations \eref{2} -- \eref{4}. By including the effects of $\alpha_c$ and $m_s$, and using ensemble theory, the EoS is given in equation \eref{12} which is the modified form of the MIT bag EoS. To solve the Einstein field equations for the anisotropic fluid sphere, we consider the $g_{tt}$ component of the metric to be Tolman-IV type. This type of metric is chosen because it is simple and singularity free. The $g_{rr}$ metric component is evaluated by via equations \eref{23} and \eref{24}. According to equation \eref{25}, the $g_{rr}$ component of the metric depends on $a$, $m_s$, $\alpha_c$ and $B_g$. The unknowns are evaluated employing matching conditions at the stellar boundary. The stability of SQM with respect to energy per baryon ($E_B$) has been analysed for different parametric choices of $\alpha_c$ and $m_s$. The variation of $E_B$ with $B_g$ is shown in \fref{fig1}. Notably, the stability window of $B_g$ for stable strange quark matter is modified because of the inclusion of $\alpha_c$ and $m_s$ and is tabulated in \tref{tab1}. In \fref{fig2}, it is observed that $E_B$ increases with the increasing $\alpha_c$. For higher $m_s$, the range of $\alpha_c$ for the stable window is reduced. Therefore, for a stable configuration in the presence of nonzero coupling ($\alpha_c$), a smaller value of $m_s$ is more acceptable. The variation of the chemical potentials, ($\mu$ and $\mu_e$) with respect to $B_g$ are shown in \fref{fig3}, for different parametric choices of $\alpha_c$ and $m_s$. It is evident from \fref{fig3} that $\mu$ increases with $B_g$, whereas, $\mu_e$ decreases. This change in $\mu$ and $\mu_e$ is more prominent as we increase $\alpha_c$ while keeping $m_s$ fixed. It is also evident that the effect of electrons is small for higher $\alpha_c$. The variation of $\mu_e$ with $\alpha_c$ is shown in \fref{fig4}, and it is evident that $\mu_e$ becomes negative at a particular value of $\alpha_c$ for a given $m_s$ and $B_g$. This is not allowed thermodynamically. Thus, an upper bound on $\alpha_c$ ($\alpha_{c,max}$) exists for each $m_s$ and $B_g$ for which $\mu_e\geq0$. In \fref{fig5}, we graphically present the restriction on $\alpha_c$, considering the value of energy per baryon ($E_B$) in three different stability windows so that the condition $\mu_e\geq0$ must hold. For an energetically stable and thermodynamically consistent configuration, both $\alpha_c$ and $m_s$ should be small. From \fref{fig5}, it is also evident that in view of stability with lower value of $m_s$, the quarks are found to be more strongly interacting. Notably, for $m_s\rightarrow0$, $\alpha_c\leq{0.584}$ for stable SQM and for the metastable SQM $\alpha_c$ lies within the range $0.584<\alpha_c<0.7$ and for $\alpha_c>0.7$ the matter is unstable. Using the modified MIT bag EoS, we solve the TOV equation to obtain the maximum mass ($M_{max}$) and other stellar parameters are tabulated in \tref{tab2}. The variation of mass ($M$) with radius ($R$) and central density ($\rho_c$) are shown in \fref{fig6} and \fref{fig7}, respectively. The black dots on each curve indicate the maximum mass, radius and central density and beyond the points $\frac{\partial{M}}{\partial{\rho_c}}<{0}$, indicating instability according to Harrison-Zeldovich-Novikov static stability condition \cite{BKH,YBZ}. It is evident form the \tref{tab2} that for a given choice of $B_g$ and $m_s$, $M_{max}$ and $R_{max}$ slightly decrease with increasing $\alpha_c$. In \fref{fig8}, the variation of the maximum mass ($M_{max}$) with $m_s$ is shown, taking $B_g=57.55~MeV/fm^3$ and different parametric choices of $\alpha_c$. The plot shows that $M_{max}$ decreases with increasing $m_s$ and that the rate of decrease is more prominent with increasing $\alpha_c$. In this model, the maximum attainable mass is $2.01~M_{\odot}$, and the corresponding radius is $10.89~km$. Thus, in this approach, a wide range of compact stars may be studied with stable SQM present inside them.\par
 
For the physical analysis of our model, we consider different compact objects with known masses,, which are assumed to be strange stars. To study the effects of $\alpha_c$ and $m_s$ on various physical properties, we randomly select two compact objects namely, $PSR~J161-2230$ and $EXO~1745-268$. In \fref{fig9} -- \fref{fig18}, all the physical parameters relevant for the stellar configuration are studied, and it is noted that $m_s$, $\alpha_c$ and  $B_g$ have some effect on these physical entities. It is evident from \fref{fig13} and \fref{fig18} that the value of anisotropy $\Delta\;(=p_t-p_r)$ is negative within a spherical region inside the stars under consideration, and then it picks up positive values. Such nature of anisotropy increases radial stability inside compact star in hydrostatic equilibrium as proposed in \cite{pretel} . Thus, it may be concluded that strange stars consisting of interacting quarks with nonzero values of strange quark mass ($m_s\neq0$) are more stable than stars consisting of noninteracting quarks. Since for a given $m_s$ and $B_g$, $\alpha_c$ largely impacts $E_B$, we have simultaneously plotted the predicted radius and $E_B$ for the selected stars against $m_s$ by taking different parametric choices of $\alpha_c$ and $B_g$ and the variation is shown in \fref{fig19}. This plot shows that the observed radius for the stars falls within the stable range.\par 
The validity of the causality conditions inside the stellar system ($0<v_r^2\leq{1}$ and $0<v_t^2\leq{1}$) are studied graphically for the selected stars in \fref{fig20} and \fref{fig21}. It is evident that $v_r^2=\frac{1}{3}$ for the parametric choice of $\alpha_c$ throughout the stars considered here, which is expected for strange matter \cite{PB}. The energy conditions for different parametric choices of $\alpha_c$, $m_s$ and $B_g$ within the stable region are shown graphically in \fref{fig22} and \fref{fig23} for the two stars. All the energy conditions are satisfied from centre to the surface. In \tref{tab3}, we have predicted the radius of a few known compact stars, which are possibly strange stars, using suitable choices of $\alpha_c$, $m_s$ and $B_g$. It is noted that the radii predicted in this model with nonzero value of $\alpha_c$ and other parameters are in aggrement with the radii estimated from observations. Therefore, if the interactions between quarks are taken into account, the model provides better radius prediction of strange stars consisting of stable SQM. The estimated value of radius of $PSR~1614-2230$ ($9.69~km$ \cite{TG}) can be predicted for different parametric choices of $m_s$, $\alpha_c$ and $B_g$, such as, ($1$) $m_s=50~MeV$, $\alpha_c=0.0$ and $B_g=72~MeV/fm^3$ ($E_B=882.371~MeV$), which corresponds to weakly bound scenario relative to $Fe^{56}$ and ($2$) for $m_s=100~MeV$, $\alpha_c=0.3$ and $B_g=69.26~MeV/fm^3$ ($E_B=937.009~MeV$), which represents strongly bound case relative to neutrons ($E_B=939~MeV$). Similarly, in the case of $EXO~1745-268$, we note the values of $m_s$, $\alpha_c$ and $B_g$ for predicting the estimated radius of $10.5~km$ \cite{FO} as, ($1$) for $m_s=50~MeV$, $\alpha_c=0.0$ and $B_g=64.56~MeV/fm^3$ ($E_B=858.96~MeV$), which corresponds to weakly bound scenario relative to $\isotope[56]{Fe}$ and ($2$) for $m_s=100~MeV$, $\alpha_c=0.3$ and $B_g=61.89~MeV/fm^3$ ($E_B=912.39~MeV$), which represents strongly bound case relative to $\isotope[56]{Fe}$ ($E_B=930.4~MeV$). 

In \tref{tab4}, natures of the SQM inside different strange stars are shown. Interestingly, the central and surface density and central pressure are modified in the presence of interactions ($\alpha_c\neq0$) between quarks. \Tref{tab5}, shows the value of compactness and surface redshift ($Z_s$). The value of $Z_s$ lies within the range as predicted previously in the articles \cite{HAB,NS,CGB}. We have also studied the variation of the moment of inertia ($I$) with mass ($M$) for the stars, and this variation is shown in \fref{fig24}. From this plot, we list the values of $I$ in \tref{tab6} for different compact stars. In addition to causality and energy conditions, the stability of our model is also studied via the analysis of the following: ($1$) the TOV equation in hydrostatic equilibrium, ($2$) Herrera cracking condition, ($3$) the adiabatic index and ($4$) the Lagrangian change of absolute value in radial pressure at the surface of the star for different parametric choices of $\alpha_c$, $m_s$ and $B_g$, and are shown in \fref{fig25} -- \fref{fig29}. It is noted that all the stability conditions holds good inside and on the surface of the stars. Thus, the internal structure of strange stars composed of stable SQM may be studied, and useful information about such configurations may be obtained considering the presence of interactions between the quarks in the QCD formalism.\\ 
   
\ack
RR is thankful to Department of Physics, Coochbehar Panchanan Barma University, for providing necessary help to carry out the research work. KBG is thankful to CSIR for providing the fellowship vide no. 09/1219(0004)/2019-EMR-I. PKC gratefully acknowledges support from IUCAA, Pune, India, under the Visiting Associateship Programme.

\end{document}